\documentclass[aps,prl,twocolumn,longbibliography]{revtex4-2}
\usepackage{graphicx}
\usepackage{subfigure}
\usepackage{adjustbox}
\usepackage{bm}
\usepackage{bbm}
\usepackage{color}
\usepackage{braket}
\usepackage{standalone}
\usepackage{multirow}
\usepackage{tikz}
\usepackage{mathrsfs}
\usepackage{dsfont}
\usepackage[colorlinks,bookmarks=true,citecolor=blue,linkcolor=blue,urlcolor=blue]{hyperref}
\usepackage{comment}
\usepackage{mathtools}
\usepackage{soul}
\usepackage{orcidlink}
\usepackage{tabularx}
\usepackage{slashed}
\usepackage{amsmath}
\usepackage{cleveref}
\usepackage{amsfonts}
\usepackage{times}
\usepackage{soul}
\usepackage{float}
\usepackage{amssymb}

\AtBeginDocument{%
   ~\newwrite\bibnotes
   ~\def\bibnotesext{Notes.bib}
   ~\immediate\openout\bibnotes=\jobname\bibnotesext
   ~\immediate\write\bibnotes{@CONTROL{REVTEX41Control}}
   ~\immediate\write\bibnotes{@CONTROL{%
    apsrev42Control,author="08",editor="1",pages="1",title="0",year="1"}}
    ~\if@filesw
    ~\immediate\write\@auxout{\string\citation{apsrev42Control}}%
   ~\fi
}%

\def\re{\text{Re}}
\def\im{\text{Im}}

\def\sgn{\text{sgn}}
 \Crefname{equation}{Eq.}{Eqs.}
\Crefname{figure}{Fig.}{Figs.}
\Crefname{section}{Sec.}{Secs.}

\def\u{\underline}

\begin{document}

\title{Non-Bloch Topology of Frustrated Non-Hermitian Flat Bands}

\author{Fan Yang\orcidlink{0000-0001-8754-7882}}
\email{fan.yang@fysik.su.se}
\author{Jiong-Hao Wang\orcidlink{https://orcid.org/0000-0002-4161-9614}}
\author{Hui Liu\orcidlink{0009-0009-4988-9561}}
\author{Emil J. Bergholtz\orcidlink{0000-0002-9739-2930}}
\email{emil.bergholtz@fysik.su.se}
\affiliation{Department of Physics, Stockholm University, SE-106 91 Stockholm, Sweden}

\date{\today}

\begin{abstract}
Frustrated lattices with exactly flat bands and compact localized states provide an ideal setting to uncover the topology of flat bands, leading to intriguing physics in two dimensions. Extending from Hermitian to non-Hermitian systems, however, band structures are drastically changed by the non-Hermitian skin effect, which is less well understood beyond one dimension. Here, we construct a broad class of two-dimensional frustrated non-Hermitian lattice models with exactly flat bands which can be naturally described in a non-Bloch band theory. Using non-Hermitian breathing Kagome and checkerboard models as illustrations, we reveal how flat bands acquire a basis-independent skin effect induced by the other bands; in the breathing Kagome model it appears exactly when the flat band enters the area spanned by the periodic-boundary spectra of the dispersive bands. A flat band becomes singular when its eigenvectors hold immovable discontinuities at band crossings. Such singularity in non-Hermitian systems is sensitive to boundary conditions and unprotected by band flatness. Non-Bloch topology underlying is uncovered in a bulk-boundary correspondence with the formation of a singular band touching in the bulk and a robust flat band skin mode on the boundary. We present a geometric approach to construct this boundary skin mode from a macroscopic sum of compact localized skin modes and trace its topological origin to two linearly independent noncontractible loop states on a non-Bloch torus. To characterize the correspondence, an invariant is proposed based on biorthogonal polarization of the zero-energy corner mode. Breaking exact flatness, we further realize nearly flat non-Bloch Chern bands.
\end{abstract}

\maketitle

\paragraph{Introduction.---}
Frustrated lattices with exactly flat bands~\cite{leykam2018a,danieli2026} provide an ideal platform to explore strongly correlated physics in 2D~\cite{bergholtz2013,derzhko2015}, due to the dominance of interaction on dispersionless energy bands. Prototypical examples including  Lieb, Kagome and checkerboard lattices have been precisely fabricated and widely applied using photonic waveguides~\cite{leykam2018p,poblete2021,danieli2024}. A topological description of these models is present at band crossings. Under periodic boundary conditions (PBC), an isolated exactly flat band can be expanded by a complete basis of compact localized states (CLSs),  of which the strictly local nature  forbids a nonzero Chern number~\cite{chen2014,li2024}. When the flat band touches other dispersive bands, immovable singularity can appear making the basis of CLSs incomplete,  a phenomenon only possible beyond 1D~\cite{bergman2008,rhim2019,rhim2021}. The flat band then turns singular with the formation of two linearly independent noncontractible loop states (NLSs) on a torus, leading to a robust boundary mode in open geometries. It establishes a bulk-boundary correspondence for singular Hermitian flat bands. 

Introducing non-Hermiticity, however, band structures under open boundary conditions (OBC)  are drastically changed by the non-Hermitian skin effect (NHSE)~\cite{hatano1996,hatano1997,zhang2022r,lee2023}. In a broad sense, it  refers to the localization of eigenmodes towards boundaries of the system driven by non-Hermiticity. The skin effect is best understood in 1D. For instance, from the PBC spectra of one-dimensional single-band models,  it can be predicted by a nonzero winding of energy on the complex plane~\cite{gong18,chen2020}. Whereas, changes in OBC spectra  and exponentially localized skin modes can be reformulated through imaginary momentum shifts in a generalized Brillouin zone (GBZ)~\cite{yao2018,flore2019}. The ensuing non-Bloch band theory is applicable in multi-band models and in certain cases  can be generalized to arbitrary dimensions~\cite{wang2024,yang2025} with well-defined topological invariants~\cite{yao2018,wang2018n,flore2019,wang2024,yang2025}. Still, beyond 1D
 the NHSE remains challenging to investigate, e.g. with momentum-dependent  GBZ~\cite{wang2024}, reliance on lattice geometries~\cite{chen2022u,hu2026} and appearance of algebraic power-law localization behaviors~\cite{kai2025}. 
So far, for NH flat bands, the relations between the skin effect, topological properties and CLSs have been addressed in one dimension~\cite{leykam2017,flore2024,dutta2025,ma2026}. 
Efforts in 2D are mainly focused on periodic lattices, including topological invariants at band crossings~\cite{oh2026} and the development of exceptional points  with degenerate eigenvectors~\cite{esparza2026}. 
To fill the gaps, here we ask following questions:
How does non-Bloch topology manifest in higher-dimensional frustrated NH flat bands? What is the role of singular flat band topology in the presence of NHSE?

 \begin{figure*}[!t]
\centering
\includegraphics[width=2\columnwidth]{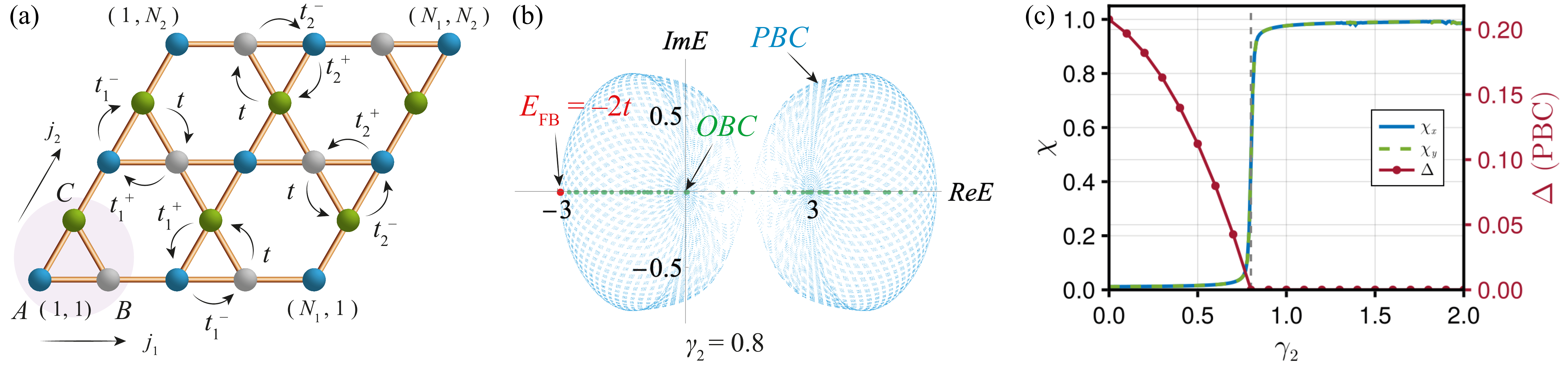}
\caption{{\bf \textcolor{black}{NH breathing Kagome model with an exactly flat band}}.  (a) NH breathing kagome lattice on a rhombus geometry with reciprocal ($t$) and nonreciprocal ($t_i^\pm  = t_i \pm \gamma_i$, $i=1,2$) hoppings. The unit cell in purple labeled by $\vec{j}=(j_1, j_2)$ is composed of three motifs $A$, $B$, $C$ in blue, gray and green.  (b) Projection  of two dispersive bands to the complex energy plane under PBC (blue) and OBC (green), where the system size is set to $N_1 = N_2 = 100$ and $5$. We choose parameters $t_1 = 1, t_2 = 2, \gamma_1 =0.2, \gamma_2 = 0.8$ and $t > 0$ fulfilling the flat band condition of \Cref{eq:fb}. The third band (red) becomes exactly flat at $E_{\text{FB}} = -2t$ regardless of boundary conditions.  The associated non-Bloch and Bloch spectra as a function of momentum $\vec{k}$ are shown in \Cref{fig:sbt}~(b) and (c).  (c) Basis-independent flat band skin effect revealed by Green's function indicator $\chi = 1$. We show indicators $\chi_x,\chi_y$ (left axis) and distance of the flat band to the dispersive bands  on the complex-energy plane $\Delta$ (right axis) as a function of $\gamma_2$. Other model parameters are kept the same as in (b).
}
\label{fig:sketch}
\end{figure*}

\begin{figure*}[!htb]
\centering
\includegraphics[width=2\columnwidth]{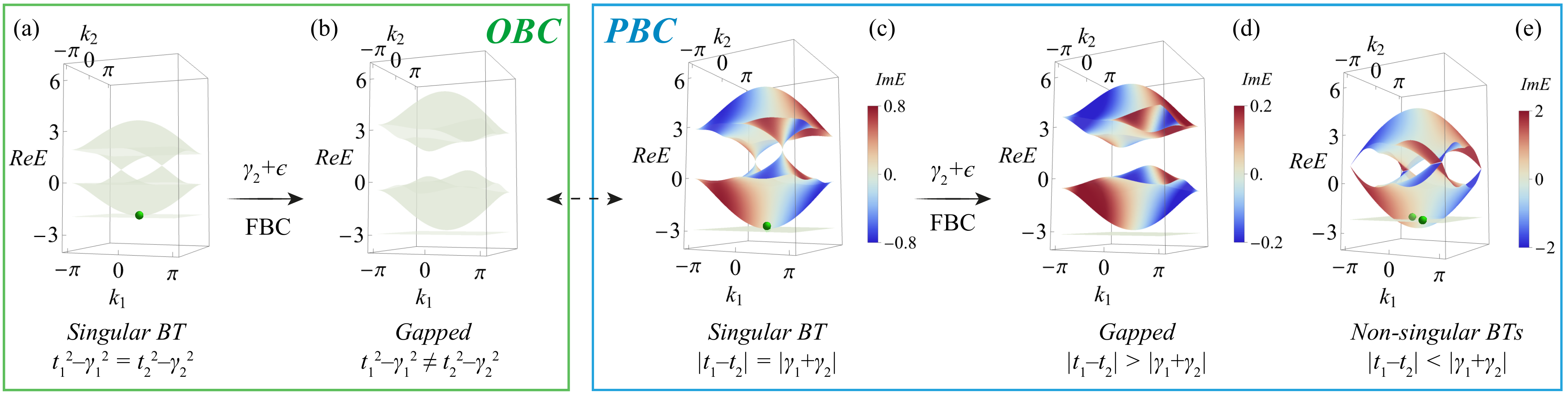}
\caption{{\bf \textcolor{black}{Extreme boundary sensitivity of singular  band touchings}}. For the NH breathing Kagome model in \Cref{fig:sketch},  under PBC a singular band touching is formed at $\vec{k} = (0, 0)$ along the parameter line $|t_1 - t_2| = |\gamma_1 + \gamma_2|$. By changing one of local parameters  away from the band touching point ($\gamma_2 \to \gamma_2 + \epsilon$) and keeping the flat band condition (FBC) through extra degree of freedom [$t$ in \Cref{eq:fb}],
the NH flat band ($E_{\text{FB}} = -2t$) can be shifted upward or downward without distortion. It leads to  a complex gap opening ($\epsilon = -0.8, |t_1 - t_2| > |\gamma_1 + \gamma_2|$) or a splitting of one singular band touching point to a pair of non-singular touching points ($\epsilon = 0.8, |t_1 - t_2| < |\gamma_1 + \gamma_2|$). Extremely sensitive to boundary conditions,  the singular band touching disappears by changing from PBC to OBC, and reappears along a different parameter line $t_1^2 - \gamma_1^2 = t_2^2 - \gamma_2^2$, which corresponds to  $\gamma_2 \simeq 1.74$ and $t > 0$ determined by \Cref{eq:fb}. A similar band shifting procedure generically opens the OBC gap.}
\label{fig:sbt}
\end{figure*}

\begin{figure*}[!htb]
\centering
\includegraphics[width=2\columnwidth]{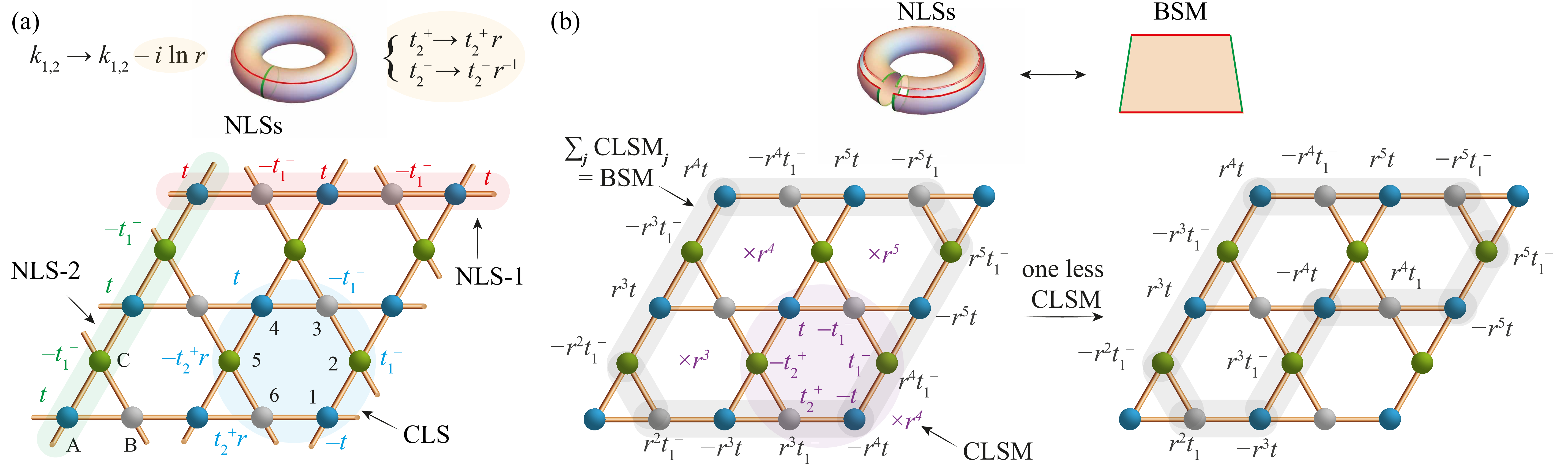}
\caption{{\bf Singular NH flat band bulk-boundary correspondence}. (a) For the NH breathing Kagome model, performing imaginary momentum shifts in the GBZ [\Cref{eq:knb}] is equivalent to modifying a pair of nonreciprocal hoppings on a torus : $t_2^\pm \to t_2^\pm r^{\pm}$, thus a non-Bloch torus is formed. When the singular band touching occurs under OBC [\Cref{fig:sbt}~(a)], the system witnesses $(N+1)$ degenerate states at the flat band energy $E_{\text{FB}}  = -2t$, where $N = (N_1-1)(N_2-1)$ is equal to the number of hexagonal plaquettes. On the non-Bloch torus, they are composed of $(N-1)$ independent CLSs (blue) and two NLSs (red and green).   (b) By cutting the torus in two directions and forming an OBC lattice, two NLSs are transformed to one BSM (gray), which exactly vanishes in the bulk. The hexagonal plaquettes support $N$ independent CLSMs (purple), the macroscopic sum of which is equal to the BSM. To reveal its robustness, one can arbitrarily
subtract a finite number of CLSMs from the sum. The operation does not destroy but only deforms the shape of the BSM.}
\label{fig:bbc}
\end{figure*}

We start by constructing a class of frustrated nonreciprocal flat band models in 2D, of which the OBC and PBC spectra are related to each other by uniform imaginary momentum shifts~\cite{yao2018,flore2019}. This spectral discrepancy is encapsulated in a non-Bloch Hamiltonian under OBC: 
\begin{gather}
    H_{\text{non-Bloch}}  = H_{\text{Bloch}} (k_1 - i\ln r_1, k_2 - i\ln r_2). \label{eq:bloch}
\end{gather}
The global feature that at least one Bloch phase factor $e^{i{(k_i-i\ln r_i)}}$ has a  modulus not equal to unity: $\exists |r_i| \ne 1$, characterizes the GBZ~\cite{yao2018,wang2024}. It also suggests that for all {dispersive} bands, the wavefunctions of eigenmodes are exponentially localized towards the boundary in the $x_i$ direction, displaying the NHSE.  While the spectra of dispersive bands are reshaped, an exactly flat band defined by $E(\vec{k}) = E_{\text{FB}}, \forall \vec{k}$, if present, is immune to the imaginary momentum shifts. It renders unique NH flat band phenomena, including the {basis-independent} skin effect (\Cref{fig:sketch}) and extreme sensitivity of singular band touchings  to boundary conditions (\Cref{fig:sbt}). Through the construction of a robust boundary skin mode (BSM, \Cref{fig:bbc}), we restore the bulk-boundary correspondence for singular non-Bloch flat bands,  and further propose a quantized invariant [\Cref{eq:nu}] inspired by generalized biorthogonal polarization of zero-energy corner mode~\cite{flore2018,elisabet2019n,elisabet2020,yang2025}.


\paragraph{\textcolor{black}{Frustrated NH Flat Band Models.}---}
The simplest way to realize a uniform GBZ is to find a similarity transformation under OBC that maps the real-space  nonreciprocal Hamiltonian $\mathcal{H}$ onto its Hermitian counterpart $\tilde{\mathcal{H}}$:
\begin{gather}
    S^{-1} {\mathcal{H}} S = \tilde{\mathcal{H}}, \quad \tilde{\mathcal{H}}^\dagger  = \tilde{\mathcal{H}}. \label{eq:s1}
\end{gather}
 $\dagger$ denotes the conjugate transpose. We can relax the condition such that the similarity transformation produces a reciprocal Hamiltonian $\tilde{\mathcal{H}}$,  of which the OBC and PBC spectra coincide in the thermodynamical limit: 
\begin{gather}
    S^{-1} {\mathcal{H}} S = \tilde{\mathcal{H}}, \quad \tilde{\mathcal{H}}^T  = \tilde{\mathcal{H}}, \notag \\
    \tilde{E}^{\text{OBC}}(\vec{k}) = \tilde{E}^{\text{PBC}}(\vec{k}), \quad N \gg 1. \label{eq:s2}
\end{gather}
 $T$ denotes the transpose. In this case, although  $\tilde{\mathcal{H}}$ can be non-Hermitian,
it hosts a normal BZ free from the NHSE.
We find such possibilities satisfying \Cref{eq:s1} or \Cref{eq:s2} are abundant in 2D frustrated NH flat band models, including modified Lieb \cite{yang2025}, breathing Kagome (main text) and modified checkerboard (End Matter) lattices. To validate the exact GBZ structure of \Cref{eq:bloch}, the Amoeba formulation \cite{ronkin1974,gelfand1994,forsberg2000,passare2004,wang2024} is provided as an alternative proof in Supplementary Material~\cite{supple}.

We first present results of  NH breathing Kagome model. Under OBC,  the lattice shown in \Cref{fig:sketch}~(a) takes the rhombus shape with  $N = (N_1-1)(N_2-1)$ hexagonal plaquettes. We denote the tight-binding Hamiltonian as $\mathcal{H} = \sum_{\vec{j}, \vec{j'},\lambda, \lambda'}  H_{(\vec{j},\lambda),(\vec{j}',\lambda')}  c^\dagger_{\vec{j},\lambda} c_{\vec{j'},\lambda'}$, where $c^\dagger_{\vec{j},\lambda}$ ($c_{\vec{j},\lambda}$) creates (annihilates) a particle on the $\lambda \in \{A, B, C \}$ motifs of the $\vec{j}$-th unit cell. There are four nonreciprocal hopping terms ($i=1,2$): $t_i^\pm = t_i \pm \gamma_i \in \mathbb{R}$ and one reciprocal hopping term $t \in \mathbb{R}$.
We identify a nonunitary gauge transform $S$, under which creation and annihilation operators are transformed according to $c_{\vec{j},A} \to r^{j_1 + j_2} \  c_{\vec{j},A}$, $c_{\vec{j},B(C)} \to \delta r \times r^{j_1 + j_2} \ c_{\vec{j},B(C)}$ and $c^\dagger_{\vec{j},A} \to r^{-(j_1 + j_2)} \  c^\dagger_{\vec{j},A}$, $c^\dagger_{\vec{j},B(C)} \to (\delta r)^{-1} \times r^{-(j_1 + j_2)} \ c_{\vec{j},B(C)}$.
With rescaling parameters $r = \sqrt{\frac{t_1^-t_2^-}{t_1^+t_2^+}}$ and $\delta r = \sqrt{\frac{t_1^-}{t_1^+}}$, the transformed Hamiltonian $\tilde{\mathcal{H}}$ becomes reciprocal with the hoppings: $t_i^\pm \to |\tilde{t_i}| = |\sqrt{t_i^+ t_i^-}|$ and $t$ unaffected. In particular,  $\tilde{\mathcal{H}}$ becomes Hermitian if $\forall t_i^\pm > 0$ [\Cref{eq:s1}]. Otherwise, $\tilde{\mathcal{H}}$ is non-Hermitian but satisfies \Cref{eq:s2}. In the following, we assume all positive $t_i^\pm$'s.
After Fourier transform $c_{\vec{j}, \lambda} = \frac{1}{\sqrt{N}}\sum_{\vec{k}} e^{i\vec{k}\cdot\vec{j}} c_{\lambda}(\vec{k})$,
the similarity transformation induces imaginary momentum shifts in the Bloch Hamiltonian [$r_1=r_2=r$ in \Cref{eq:bloch}], rendering the non-Bloch form :
\begin{align}
       & H_K = H(k_1 - i\ln r, k_2 - i\ln r)  \label{eq:knb} \\
       = & \begin{pmatrix}
	  		0 & t_1^+ + t_2^- r^{-1}e^{-ik_1} &    t_1^+ + t_2^-r^{-1} e^{-ik_2} \\
			t_1^- + t_2^+r e^{ik_1} & 0 & t+te^{i(k_1-k_2)} \\
			 t_1^- + t_2^+ re^{ik_2} & t+te^{-i(k_1-k_2)} & 0
		    \end{pmatrix}. \notag 
	\end{align}
Here, we apply the basis ${\varphi} (\vec{k}) = (c_A (\vec{k}), c_B (\vec{k}),  c_{C} (\vec{k}))^T$ and
$k_i = \vec{k} \cdot \vec{a}_i$ are associated with two unit vectors along $j_i$ directions: $\vec{a}_1 = (1,0), \vec{a}_2 = (\frac{1}{2},\frac{\sqrt{3}}{2})$.

A flat band arises at $E_{\text{FB}} = -2t$ if
 \begin{gather}
    t_1^2 + t_2^2 - \gamma_1^2 - \gamma_2^2 = 2t^2. \label{eq:fb}
 \end{gather}
This flat band condition is invariant under OBC or PBC. One can verify through the Bloch Hamiltonian
  by setting $r=1$ in \Cref{eq:knb}. In \Cref{fig:sketch}~(b), we numerically calculate finite-size spectra of the model at the flat band condition. Switching from PBC to OBC, two complex  dispersive bands (blue) convert to real bands (green, $\tilde{\mathcal{H}}$ being Hermitian) while the flat band (red) remains untouched. 

In dispersive bands, the nonunitary gauge transform $S$ leads to skin effect. However, in the flat band, the arbitrary choices of basis due to the high degeneracy makes the skin effect ambiguous, or basis-dependent. For example, the CLSs are still strictly localized within the original confined region, despite the similarity transformation. We show the 2D flat band skin effect is basis-dependent when the flat band is separated from the other bands in the complex energy plane, while is basis-independent when the flat band is embedded within the finite area of other bands~\cite{ma2026}. We introduce a source $|s\rangle$ at the right top corner of the sample in Fig.~\ref{fig:sketch}(a) under OBC and calculate the response using the Green's function $|r\rangle=G(E)|s\rangle$, where the Green's function is $G=(E+i\epsilon-\mathcal{H})^{-1}$ with $E$ being the flat band energy and $i\epsilon$ a tiny imaginary part, independent of the basis choice. We define the indicators $\chi_\mu=\sum_i^{L_\mu}(\mu_i/L_\mu)|\langle i|r\rangle|^2$ with $\mu=x,y$, where $\mu_i$ is the position coordinate in $\mu$ direction, indexed from right (top) to left (bottom) in $x$ ($y$) direction. If $|r\rangle$ is localized on the left (bottom) boundary, $\chi_x$ ($\chi_y$) will be 1, exhibiting the skin effect. Otherwise, if $|r\rangle$ is localized on the right (top) boundary as the source $|s\rangle$, $\chi_x$ ($\chi_y$) is 0, indicating the absence of skin effect in the response. As shown in \Cref{fig:sketch}~(c),  $\chi_x$ and $\chi_y$ coincide and jump from 0 to 1 at $\gamma_2=0.8$, as indicated by the vertical dashed line, which is just the band touching point under PBC as shown by the distance of the flat band to the other bands $\Delta$ under PBC, consistent with \Cref{fig:sketch}~(b) or \Cref{fig:sbt}~(c). Therefore, the flat band exhibits basis-independent skin effect when embedded in the interior of dispersive bands under PBC, as shown in \Cref{fig:sbt}~(e), while when the flat band is isolated as in \Cref{fig:sbt}~(d), the basis-independent skin effect is absent.

\paragraph{Singular Flat Band Touchings.---} In 2D,  topologically nontrivial NH flat band touchings share three main features: boundary sensitive, singular and not protected by  band flatness.

Boundary sensitivity has its origin in non-Hermiticity. Let us look at  band touchings (BTs) in the breathing Kagome model (\Cref{fig:sbt}). From the non-Bloch spectra, two dispersive bands read $E_\pm(\vec{k}) = t \pm \sqrt{3t^2 + q_{\vec{k}}}$ with $q_{\vec{k}} = 2t^2 \cos(k_1 - k_2) + t_1^-t_2^-r^{-1}(e^{-ik_1}+ e^{-ik_2}) + t_1^+t_2^+r(e^{ik_1}+ e^{ik_2})$. Once $q_{\vec{k}} = 6t^2$, a band touching appears
between the lower band  $E_-$ and the flat band $E_{\text{FB}} = -2t$. Same argument applies to the Bloch spectra by taking $q_{\vec{k}} = q_{\vec{k}}|_{r=1}$. Yet, sensitive to boundary conditions, the band touching  formed at  $\vec{k} = (0, 0) = \vec{0}$ is reached along different parameter lines:
  \begin{align}
      &\text{OBC:} \quad t_1^2 - \gamma_1^2 =  t_2^2 - \gamma_2^2, \label{eq:sk0} \\
     &\text{PBC:} \quad  |t_1 - t_2| = |\gamma_1 + \gamma_2|. \notag
  \end{align}
Here, we have applied the flat band condition [\Cref{eq:fb}] to release the degree of freedom in $t$.
In the Hermitian limit $\vec{\gamma} = (\gamma_1, \gamma_2) = \vec{0}$, provided $\forall t^\pm_i > 0$, the two band touching lines converge to one: $t_1 = t_2 = t$ which reduces the model to a Kagome metal~\cite{rhim2019}.

For singularity, we classify a band touching  to be singular if it induces immovable discontinuities in normalized flat band  eigenvectors. NH Hamiltonians host two sets of eigenmodes ($m$), the left and right ones: $\mathcal{H} |{{\psi}}_{R,m}\rangle = E_m |{{\psi}}_{R,m}\rangle$, $\mathcal{H}^\dagger |{{\psi}}_{L,m}\rangle  = E_m^* |{{\psi}}_{L,m}\rangle$, where $*$ denotes the complex conjugate. Adopting the biorthogonal normalization $\langle {\psi}_{L,m}  | {\psi}_{R,m'} \rangle = \delta_{m,m'}$, an eigenvalue decomposition of NH Hamiltonians is achieved: $\mathcal{H} = \sum_m E_m |{\psi}_{R,m}\rangle \langle {\psi}_{L,m}|$.
For the breathing Kagome flat band, the right non-Bloch eigenvector takes the form:
  \begin{gather}
      {v}_{R} (\vec{k}) = \eta_{\vec{k}} \begin{pmatrix}
          t - te^{-i(k_1 - k_2)} \\
          -t_1^- + t_2^+ re^{ik_2} \\
          t_1^- e^{-i(k_1-k_2)} -  t_2^+ re^{ik_2} 
      \end{pmatrix}. \label{eq:kbe}
  \end{gather}
From the symmetry $H_K^T (\vec{k}, \vec{\gamma}) = H_K(-\vec{k}, -\vec{\gamma})$, the left eigenvector ensues: $v_{L}^*(\vec{k},\vec{\gamma}) = v_{R}(-\vec{k},-\vec{\gamma})$. If one chooses a biorthogonal basis $  {v}^*_{L} (\vec{k}) \cdot  {v}_{R} (\vec{k}) = 1$, the normalization factor follows:  $\eta_{\vec{k}} = 1/\sqrt{6t^2 - q_{\vec{k}}}$. At the band touching with $q_{\vec{0}} = 6t^2$, $\eta_{\vec{k} \to \vec{0}} \to \infty$. Whereas, all components of  $v_{R}$ and $v_{L}$ vanish according to the band touching condition [\Cref{eq:sk0}]. The singular nature of band touching is revealed in the fact that when one approaches the touching point, the value of $v_{R/L}$ depends on which route is taken, e.g. $\lim_{\delta \to 0} v_R(\delta, 0) \ne \lim_{\delta \to 0} v_R(0,\delta)$. Since the band touching  is gauge-independent, such discontinuity is immovable by another gauge choice of $v_{R/L}$.  
A closer scrutiny also shows that singularity does not rely on the way of normalization. In the current form of \Cref{eq:kbe}: $v_R = \eta_{\vec{k}} (w_A , w_B, w_C)^T$, the components $w_\lambda$ have no $\vec{k}$-dependent common factor and each  is vanishing at the band touching. If we normalize the right-right eigenvectors $ {v}^*_{R} (\vec{k}) \cdot  {v}_{R} (\vec{k}) = 1$, $\eta^R_{\vec{k}} = 1/\sqrt{\sum_{\lambda} |w_\lambda|^2}$  still diverges at $\vec{k} = \vec{0}$ and $  {v}_{R}$ remains discontinuous. \textcolor{black}{As a comparison, we can examine the PBC case by taking $r \to 1$ in \Cref{eq:kbe}. Along the singular line $t_1^- = t_2^+$ ($t_1^+ = t_2^-$), the discontinuity is revealed only in one of the Bloch eigenvectors $v_R$ ($v_L$).}

Instability of singular flat band touchings is not fully explored. We unveil a type of instability caused by perturbations that preserve band flatness, not yet discussed in Hermitian systems~\cite{rhim2019}.
It can be realized by tuning parameters of the Hamiltonian away from the singular band touching condition [\Cref{eq:sk0}] while keeping the independent flat band condition [\Cref{eq:fb}].
In the NH breathing Kagome model (\Cref{fig:sbt}), we examine such instability under OBC and PBC with the singularity perturbed by $\gamma_2$ and the band flatness preserved through $t$. Interestingly, 
in the real non-Bloch spectra, $E_-$ reaches a lower bound at the singular band touching. Any perturbation that breaks the band touching  always opens up a gap, leading to a metal-to-insulator transition. In the complex Bloch spectra, on the contrary, there is no similar bound. A unique NH flat band phenomenon arises --- a singular Bloch band touching point ($|t_1-t_2| = |\gamma_1 + \gamma_2|$) can split up into a  pair of non-singular touching points ($|t_1-t_2| < |\gamma_1 + \gamma_2|$) with exact band flatness. At non-singular band touchings~\cite{note}, every component $w_\lambda$ in $v_R$ is non-zero. We are able to remove the divergence in $\eta_{\vec{k}}$ or the discontinuity in $v_R$ by normalizing the right-right eigenvectors. 
In the Hermitian limit ($\gamma_1, \gamma_2 \to 0$), similar perturbation  that preserves band flatness can only open the gap, inducing the transition from the Kagome metal ($t_1=t_2=t$) to the breathing Kagome insulator ($t_1\ne t_2, t_1^2 + t_2^2 = 2t^2$). 

Unlike Ref.~\cite{leykam2017},
 flat band touchings discussed in our models are not exceptional points.

\paragraph{Singular Non-Bloch Flat Band Topology and Boundary Skin Modes.---}
We proceed to uncover the topology of non-Bloch flat bands at singular band touchings.

 To find the BSM, we develop a geometric approach  inspired by Hermitian analogues~\cite{rhim2019}. We start from  the state counting of the singular flat band on a non-Bloch torus [NBT, \Cref{fig:bbc}~(a)], followed by cutting the torus in two directions and reconstructing all states under OBC [\Cref{fig:bbc}~(b)]. On the NBT, 
 to accommodate the GBZ shift [\Cref{eq:knb}], it is natural to redefine nonreciprocal hoppings: $t_2^\pm \to t_2^\pm r^{\pm 1}$, so that the torus Hamiltonian inherits the non-Bloch spectra. 
 
 A simple way of counting states is to construct a total of $N$ CLSs, one per hexagonal plaquette. For the right set, let us make a superposition of flat band biorthogonal eigenmodes:
 \begin{gather}
       |\chi_{R} (\vec{j}) \rangle = \mathcal{N}_\chi \sum_{\vec{k} \in  \text{BZ}} \frac{1}{\eta_{\vec{k}}}  e^{-i\vec{k} \cdot \vec{j}}   {\psi}_{R} (\vec{k}), \label{eq:chr} \\
         {\psi}_{R} (\vec{k}) =  \sum_{\vec{j}'} \sum_{\lambda} g_{\vec{j}'}  e^{i\vec{k}\cdot \vec{j}'} {v}_R(\vec{k},\lambda) |\vec{j}', \lambda \rangle, \notag
 \end{gather}
with $\text{BZ} = \re[\text{GBZ}]$ and $\mathcal{N}_\chi$ the normalization factor. The superposition factor $1/\eta_{\vec{k}}$ is nonzero except at the singular band touching, $\vec{k} = \vec{0}$. $\chi_{R}$ then expands an incomplete basis composed of $(N-1)$ independent states on the NBT. Changing from NBT to OBC, the Hamiltonian recovers its original form: $t_2^\pm  r^{\pm 1} \to t_2^\pm$. We add an extra $g$ factor to account for the GBZ shift:
  \begin{gather}
   g_{\vec{j}} = \begin{cases}
     \   1, & \text{NBT};  \\
      \   r^{j_1 +  j_2}, & \text{OBC}.
   \end{cases} \label{eq:gj}
  \end{gather}
In \Cref{eq:chr}, the sum over $\vec{k}$ leads to localized states  within a finite range of the lattice: $ |\chi_{R} (\vec{j}) \rangle = \mathcal{N}_\chi \sum_{\vec{j}'} \sum_\lambda {A}_{R, \vec{j},\vec{j}',\lambda} |\vec{j}', \lambda \rangle$, where
\begin{gather}
    {A}_{R, \vec{j},\vec{j}'} =   g_{\vec{j}'} \begin{pmatrix}
          t\delta_{\vec{j}',\vec{j}} - t\delta_{\vec{j}',\vec{j}+\vec{a}_3} \\
          -t_1^-\delta_{\vec{j}',\vec{j}} + t_2^+ r\delta_{\vec{j}',\vec{j}-\vec{a}_2} \\
          t_1^- \delta_{\vec{j}',\vec{j}+\vec{a}_3} -  t_2^+ r\delta_{\vec{j}',\vec{j}-\vec{a}_2}
      \end{pmatrix},
\end{gather}
and $\vec{a}_3 = \vec{a}_1 - \vec{a}_2$. It enables us to build one CLS per   hexagonal plaquette [\Cref{fig:bbc}~(a)]. On each hexagon with sites indexed by  $u = 1,2,\dots,6$  and the top edge $(4,3)$  situated in the $\vec{j}$-th unit cell, the right CLS takes the form:
$|\chi_{R} (\vec{j}) \rangle = \mathcal{N}_\chi \sum_{u=1}^6 \u{A}_{R, \vec{j}, u} |\vec{j}, u \rangle$ with
\begin{gather}
 {A}_{R, \vec{j}}^T =   \begin{cases}
       (-t, t_1^-, -t_1^-, t, -t_2^+r,t_2^+r), &\text{NBT} \\
        (-t, t_1^-, -t_1^-, t, -t_2^+, t_2^+), &\text{PBC} \\
      r^{j_1+j_2} (-t, t_1^-, -t_1^-, t, -t_2^+, t_2^+), &\text{OBC} 
  \end{cases}   \label{eq:cls}
\end{gather}
and the left CLS follows from $|\chi^*_{L} (\vec{j},\vec{\gamma}) \rangle = |\chi_{R} (\vec{j},-\vec{\gamma}) \rangle$.
Though the right and left sets of CLSs are normalized to each other: $\langle \chi_{L} (\vec{j}) | \chi_{R} (\vec{j}) \rangle = 1$ which leads to
$\mathcal{N}_\chi = (\sqrt{6}t)^{-1}$,  they are not biorthogonal: $\langle \chi_{L} (\vec{j}\pm \vec{a}_i) | \chi_{R} (\vec{j}) \rangle \ne 0$, $\forall i \in \{1,2,3\}$. Our construction is consistent with the Hermitian limit ($\vec{\gamma} = \vec{0}$) where left and right CLSs become identical and form a non-orthogonal basis~\cite{rhim2019}. 

On the NBT [\Cref{fig:bbc}~(a)], a state counting on the right set reveals that there are only $(N-1)$  independent CLSs (blue) due to the constraint  $\sum_{\vec{j}} |\chi^{\text{NBT}}_{R} (\vec{j}) \rangle = 0$ imposed by the non-Bloch singular band touching condition [\Cref{eq:sk0}]. We identify two more independent eigen states --- a pair of \textcolor{black}{right} NLSs (red and green) circulating different directions of the torus. Together, they contribute to a total of $(N+1)$ degenerate states  along the non-Bloch singular line [\Cref{fig:sbt}~(a)].
In the non-singular case with a gap opening, $\sum_{\vec{j}} |\chi_{R}^{\text{NBT}} (\vec{j}) \rangle \ne 0$. CLSs constructed on $N$ plaquettes are independent to one another, thus  forming a complete non-Bloch basis of the flat band. Moreover, taking the limit $r \to 1$ one obtains $N$ \textcolor{black}{right} CLSs  in the PBC scenario [\Cref{eq:cls}]. The same pair of  \textcolor{black}{right} NLSs in \Cref{fig:bbc}~(a) completes the basis 
\textcolor{black}{along the Bloch singular line  $t_1^- = t_2^+$}. \textcolor{black}{For the other Bloch singular line $t_1^+ = t_2^-$, there arises a pair of left NLSs obtained by flipping $\gamma_1 \to -\gamma_1$ in the right set.}

Now, we consider a third geometry, the OBC lattice in a rhombus shape [\Cref{fig:bbc}~(b)] obtained by cutting the non-Bloch torus in $j_1$ and $j_2$ directions. CLSs on the NBT are no longer an eigen state under OBC. Recalling the similarity transformation ${\mathcal{H}}  = S\tilde{\mathcal{H}}S^{-1} = \sum_m E_m S|\tilde{\psi}_{m}\rangle \langle \tilde{\psi}_{m}|S^{-1}$ and  $S^\dagger = S$, one relates non-Hermitian and Hermitian eigenmodes: $| {\psi}_{R,m} \rangle = S |\tilde{\psi}_{m} \rangle$ and $| {\psi}_{L,m} \rangle = S^{-1} |\tilde{\psi}_{m} \rangle$. 
 It is necessary to perform imaginary momentum shifts in the wave function $\psi_R$ through the $g$ factor [\Cref{eq:chr} and \Cref{eq:gj}]. The same superposition procedure leads to compact localized skin modes [CLSMs, \Cref{eq:cls}].  Shown in \Cref{fig:bbc}~(b), on isolated hexagons CLSMs (purple) share the local structure (relative site weights) as CLSs under PBC, because plaquette states live in a local environment insensitive to boundary conditions. Yet,  the skin localization factor $|r| \ne 1$ acts in a global way such that the sum of $N$ independent CLSMs amounts to a BSM (gray): 
    \begin{gather}
    |\Psi_{R,\text{boundary}} \rangle = \sum_{\vec{j}} |\chi_{R}^{\text{OBC}} (\vec{j}) \rangle. \label{eq:bsm}
\end{gather}
Along the non-Bloch singular line, the BSM  is exactly vanishing in the bulk whereas shows the NHSE on the boundary of the 2D lattice [\Cref{fig:bbc}~(b), bottom left]. Its topological origin  can be traced back to the pair of NLSs through the deformation of the non-Bloch torus [\Cref{fig:bbc}~(b), top]. Moreover, the BSM is robust against plaquette defects and changes of the lattice shape: taking away an arbitrary finite number of CLSMs from the macroscopic sum in \Cref{eq:bsm}, the resulting $\Psi'_{R,\text{boundary}}$ remains an eigenmode [\Cref{fig:bbc}~(b), bottom right].
In comparison, the sum of all CLSs under PBC behaves differently: $\sum_{\vec{j}} |\chi_{R}^{\text{PBC}} (\vec{j}) \rangle$  vanishes  \textcolor{black}{along the Bloch singular line  $t_1^- = t_2^+$} and leads to a bulk mode elsewhere. \textcolor{black}{The other Bloch singular line $t_1^+ = t_2^-$ is signified by $\sum_{\vec{j}} |\chi_{L}^{\text{PBC}} (\vec{j}) \rangle = 0$.}
 
\paragraph{Biorthogonal Invariant.---} Notably, the NH breathing Kagome model holds an exact zero-energy corner mode occupying  the motif $A$ only~\cite{elisabet2019n}: 
\begin{gather}
   |{\psi}_{R/L, 0} \rangle = \mathcal{N}_{R/L} \sum_{\vec{j}}  (r_{R/L})^{j_1+j_2} c^\dagger_{ \vec{j}, A} |0\rangle, \label{eq:corner}
 \end{gather}
 where $r_R = -t_1^-/t_2^+$, $r^*_L = -t_1^+/t_2^-$ and $\mathcal{N}_{R/L}$ is determined by  $\langle \psi_{L,0}  | \psi_{R,0}  \rangle = 1$. Along the non-Bloch singular line or equivalently $r_Rr^*_L=1$, the biorthogonal polarization of the corner mode is half-quantized  $\vec{P} = (P_1, P_2) = (1/2, 1/2)$:
 \begin{gather}
    P_l =  1- \lim_{N_l \to \infty} \left< \psi_{L,0}\left| \frac{\sum_{\vec{j},\lambda}  j_l \hat{\Pi} (\vec{j}, \lambda) }{N_l} \right| \psi_{R,0}\right>, \label{eq:pol}
 \end{gather}
with $\hat{\Pi} (\vec{j}, \lambda) = |\vec{j}, \lambda \rangle \langle \vec{j}, \lambda|$  the density operator. 
 Whereas for the non-singular case with a gap opening, $\vec{P} = (0, 0)$ or $(1,1)$ when  $r_Rr^*_L > 1$ or $0 < r_Rr^*_L < 1$.
 
Hence,  we can construct a biorthogonal invariant $\nu$  based on the polarization of the corner mode:
\begin{gather}
    \nu = \Pi_{l=1}^d \sin(\pi P_l). \label{eq:nu}
\end{gather}
$\nu= (0) \ 1$ for (non-)singular non-Bloch flat bands. Here, $d \ge 2$ and $\nu$ can be applied to $d$-dimensional nonreciprocal hypercubic models with  $(d-1)$ degenerate NH flat bands and exact corner modes \cite{yang2025}. In the latter case, flat bands become non-singular $\nu = 0$ if at least one component of polarization is integer-valued: $\exists P_l = 0$ or $1$.  It can be checked that our biorthogonal invariant $\nu$ applies to the Hermitian limit ($\gamma_1, \gamma_2 \to 0$) as well.

Across the phase transition of current model [\Cref{fig:sbt}~(a) and (b)], the BSM at $E_\text{FB} = -2t$  and the corner mode at $E_0 = 0$  can be viewed as two competing boundary modes. Cast in the biorthogonal density,
  \begin{gather}
      n_{L R, m}(\vec{j}, \lambda) =  \frac{\langle {\psi}_{L,m}| \hat{\Pi} (\vec{j}, \lambda) | {\psi}_{R,m} \rangle }{\langle {\psi}_{L,m} | {\psi}_{R,m} \rangle}, \label{eq:bd} 
  \end{gather}
 for  (non-)singular non-Bloch flat bands with $r_Rr^*_L \  (\ne) =1$, it is the (corner) boundary skin mode that faithfully reveals a boundary-localized spatial structure.

\begin{figure}[!t]
\centering
\includegraphics[width=\columnwidth]{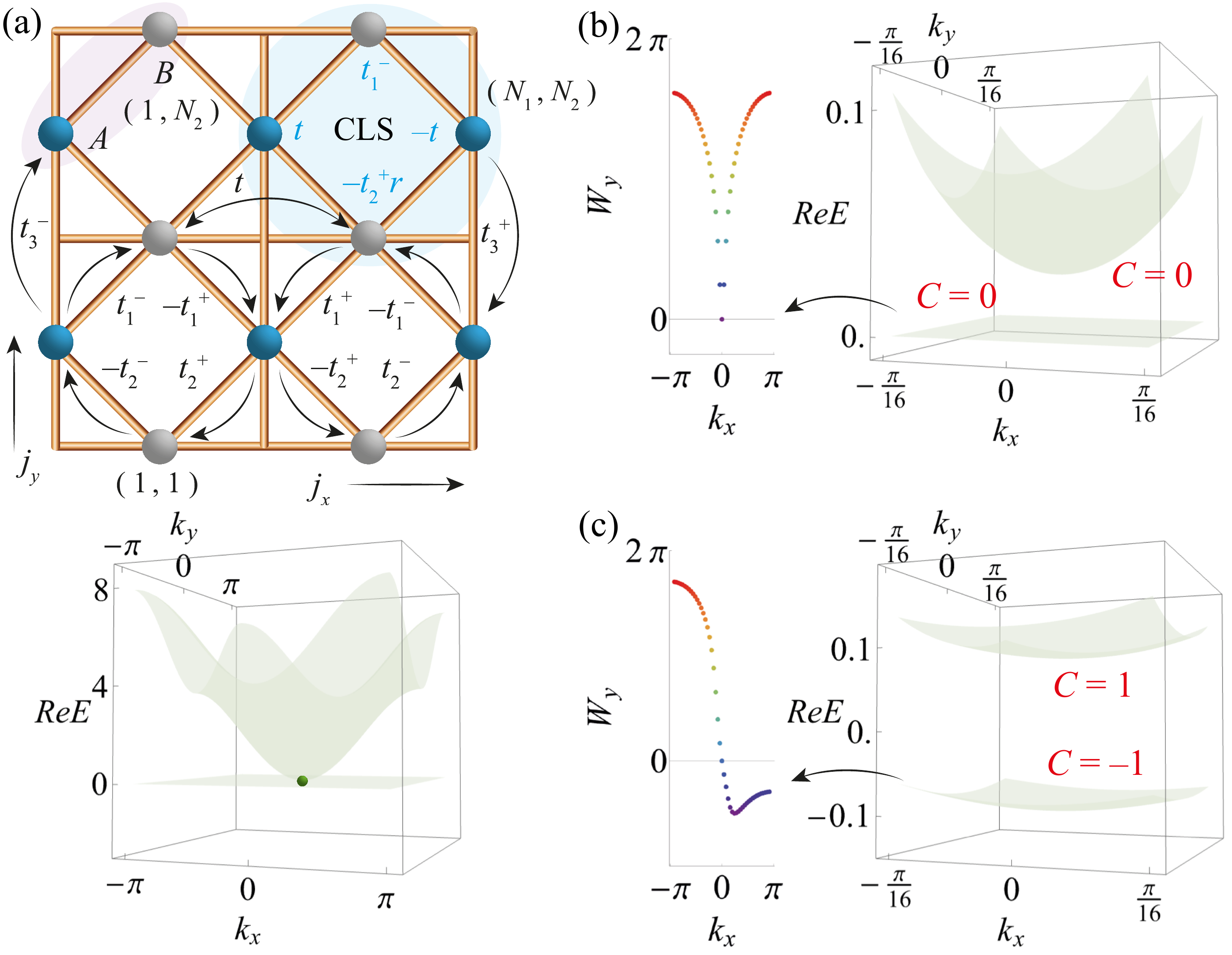}
\caption{{\bf \textcolor{black}{No-go theorem in non-Bloch flat bands}}. (a) NH checkerboard lattice on a square geometry with reciprocal ($t$) and nonreciprocal ($t_i^\pm  = t_i \pm \gamma_i$, $i=1,2,3$) hoppings. 
The unit cell in purple labeled by $\vec{j}=(j_x, j_y)$ is composed of two motifs $A$ and $B$ in blue and gray. By tuning $t_3^\pm  = ({ t_2^\pm}/{t_1^\mp})t$ and assigning mass terms $m_A = - [1+(t_2^+t_2^-)/(t_1^+t_1^-)]t$, $m_B = -2t$, we obtain the non-Bloch Hamiltonian in \Cref{eq:cnbh}. When $t_1 = 1$, $t_2 = 2$, $\gamma_1 = 0$, $\gamma_2  = \sqrt{3}$ and $t = -1$,  the  non-Bloch spectrum becomes real and  hosts
a flat band at  $E_{\text{FB}} = 0$, which displays a singular band touching with the dispersive band. 
(b)-(c) Under perturbations, a gap can be opened. The flat band either remains exactly flat and topologically trivial or becomes dispersive with the possibility of acquiring a nonzero non-Bloch Chern number. In (b), we vary $\gamma_i \to \gamma_i + \epsilon  \ (i = 1,2)$ and keep the flat band condition by tuning $t = -\sqrt{t_1^2 - \gamma_1^2}$. The net zero change of Berry phase  confirms the no-go theorem. In (c), we add a perturbation $\delta H = \epsilon \sigma_y$ to break the exact flatness. Microscopically, it can  be done by turning on imaginary hoppings inside each unit cell. 
A topological nearly flat band is created with Berry phase undergoing a  change of $(-2\pi)$.  In both cases, we set $\epsilon = -0.1$.}
\label{fig:chern}
\end{figure}

\paragraph{Discussion and Outlook.---} To summarize, by constructing a class of 2D frustrated NH flat band models with a non-Bloch description, 
we extend the classification of singular and non-singular flat bands from  the Hermitian~\cite{rhim2019} to non-Hermitian regime and from periodic lattices~\cite{oh2026} to open geometries with concurrent basis-independent skin effect. Moreover, in the breathing Kagome model an exactly flat band displays a basis-independent skin effect precisely when it lies inside the PBC spectral area of the dispersive bands; this is the 2D counterpart of the 1D point-gap criterion of Ref.~\cite{ma2026}.
In End Matter, we go on to discuss the scenario with broken band flatness. We show how to realize nearly flat non-Bloch Chern bands (\Cref{fig:chern}) using the example of  NH modified checkerboard model.


 In the future, experimental realizations of frustrated non-Bloch flat bands can be pursued in photonic lattices~\cite{el2019,photonicNHBBC}, circuits~\cite{helbig2020generalized,Neupert_2020}, lasing \cite{amelio2024}, and mechanical metamaterials~\cite{Brandenbourger_2019,Ghatak2020,Ma2022,veenstra2023,ma2026}. In photonic crystals, one can prepare the boundary skin mode with an input beam and test the nondiffracting property. On the other hand, to enrich the theory, singular non-Bloch band touchings also occur among more than two bands in NH modified Lieb model~\cite{yang2025}. Apart from NLSs and CLSs, we expect higher degeneracy at the singularity to support extended states on a non-Bloch torus. They become extended skin modes under OBC. 
By dissipative state preparation~\cite{schomerus2013,emil2023,meng2023,yang2025q}, it is possible to observe Liouvillian skin effects~\cite{fei2019,ueda2021,yang2022,kohei2023,halati2026} from boundary and extended skin modes, leaving dynamical footprints of singular non-Bloch flat bands. 
For Hermitian flat bands, the pair of NLSs can be linearly dependent, which helps remove the band touching singularity and enforces nontrivial invariants in 2D and 3D~\cite{liu2026}. Similar condition may be considered in the non-Hermitian case.
Introducing interactions in our models, it is promising to explore NH topological orders~\cite{hao2025}. As to the influence of disorder~\cite{danieli2026}, by analogy to 1D~\cite{leykam2017}, NH flat band induced CLSs in 2D are expected to be robust against weak disorder.

\paragraph{Acknowledgements.---} We thank 
Rodrigo Arouca,  J.~Lukas K.~König, Guancong Ma and Kang Yang for discussions. 
This work was supported by the Knut and Alice Wallenberg Foundation (KAW) via the Wallenberg Academy Scholars program (2023.0256), the project Dynamic Quantum Matter (2019.0068) and Wallenberg Scholars, the Swedish Research Council under the VR Starting Grant 2024-05213, and the G\"oran Gustafsson Foundation for Research in Natural Sciences and Medicine.

\bibliography{sample}

\begingroup
\onecolumngrid
\section{{E\MakeLowercase{nd} M\MakeLowercase{atter}}}
\endgroup
\twocolumngrid



\paragraph{Nearly Flat Non-Bloch Chern Bands.---} Here, we explore conditions to realize non-Bloch Chern bands in frustrated NH flat band models.  

If we keep exact band flatness, as has been shown, any gapped thus non-singular NH flat band can be expanded by a complete basis of CLSs on a non-Bloch torus. In Hermitian and non-Hermitian periodic lattices~\cite{chen2014,li2024,oh2026}, such  strictly local band structure forbids a nonzero  Chern number. In the following, we generalize this no-go theorem to the non-Bloch case with the example of NH modified checkerboard model.

In \Cref{fig:chern}~(a), we place the checkerboard lattice on a square geometry under OBC. Each unit cell (purple) is composed of two sites $A, B$ at the position $\vec{j}=(j_x, j_y)$ with unit vectors $\vec{a}_x = (1,0)$ and $\vec{a}_y = (0,1)$. 
Among microscopic parameters, we set the reciprocal hopping ($t$) and nonreciprocal hoppings ($t_1^\pm, t_2^\pm$) as free parameters while fix the next-nearest-neighbor nonreciprocal hopping and two mass terms:  $t_3^\pm  = ({ t_2^\pm}/{t_1^\mp})t$, $m_A = - [1+(t_2^+t_2^-)/(t_1^+t_1^-)]t$, $m_B = -2t$. We assume $t_i^\pm > 0, \forall i=1,2$. It allows us to find  a similarity transformation towards a Hermitian Hamiltonian [\Cref{eq:s1}]: $c_{\vec{j},A} \to r^{j_y} \  c_{\vec{j},A}$, $c_{\vec{j},B} \to \delta r \times r^{j_y} \ c_{\vec{j},B}$ and $c^\dagger_{\vec{j},A} \to r^{-j_y} \  c^\dagger_{\vec{j},A}$, $c^\dagger_{\vec{j},B} \to (\delta r)^{-1} \times r^{- j_y} \ c_{\vec{j},B}$ 
with rescaling parameters $r = \sqrt{\frac{t_1^-t_2^-}{t_1^+t_2^+}}$ and $\delta r = \sqrt{\frac{t_1^-}{t_1^+}}$.  In the basis ${\varphi} (\vec{k}) = (c_A (\vec{k}), c_B (\vec{k}))^T$,
the non-Bloch Hamiltonian of the checkerboard model takes the form:
\begin{gather}
     H_C = H(k_x, k_y - i \ln r) =  \begin{pmatrix}
        -t \omega_y^+ \omega_y^- & t_1^+ \omega_y^- \omega_x^*  \\
         t_1^-\omega_y^+ \omega_x &  -t \omega_x \omega_x^*
    \end{pmatrix}, \notag \\
      \omega_x = 1-e^{ik_x}, \quad \omega^{\pm}_y = 1-\frac{t_2^\pm r^{\pm 1}}{t_1^\mp}e^{\pm ik_y}.\label{eq:cnbh}
\end{gather}
Under the flat band condition:
$t_1^2 - \gamma_1^2 = t^2$,
one obtains $E_\text{FB} = E_1 = 0$ and $E_2 (\vec{k}) = -t[2(1-\cos k_x)+(1-\frac{t_2^+r}{t_1^-}e^{ik_y})(1-\frac{t_2^-r^{-1}}{t_1^+}e^{-ik_y})]$. Shown in \Cref{fig:chern}~(a), a singular band touching appears at $\vec{k} = (0,0)$  when $t_1^2 - \gamma_1^2 =  t_2^2 - \gamma_2^2$.

In 2D, NH Chern number~\cite{fu2018} is defined by 
\begin{gather}
   C_{\alpha \beta} = \frac{1}{2\pi i} \int_{T^2} dk_x dk_y \epsilon^{ij} \partial_{k_i} \langle {v}_{\alpha}(\vec{k})|\partial_{k_j} |  {v}_{\beta}(\vec{k}) \rangle,
\end{gather}
where $\epsilon^{xy} = \epsilon^{yx} = -1$ and $T^2 = [-\pi, \pi]^2$ denotes a two-dimensional torus.
$v_{\alpha = R,L}$ stand for right and left non-Bloch eigenvectors expanding the band. In the presence of a gap, it turns out that all four Chern numbers are equal~\cite{fu2018}: $C_{LR} = C_{RL} = C_{RR} = C_{LL} = C$. 

$H_C$ falls into the class of two-band non-Bloch Hamiltonians which can be represented by Pauli matrices $\vec{\sigma}= (\sigma_x, \sigma_y, \sigma_z)$ and $\sigma_0 = \mathbbm{1}$: $H(\vec{k}) = \vec{d}(\vec{k}) \cdot \vec{\sigma} + d_0 \sigma_0$ with $\vec{d}= (d_x, d_y, d_z)$ and $d_\lambda \in \mathbb{C}$. Energy dispersions are resolved as $E_\pm = d_0 \pm d$ with $d = \sqrt{d_x^2 + d_y^2 + d_z^2}$. For the lower band $E_-$,  the non-Bloch Chern number adopts a simplified form in $C_{LR}$~\cite{wang2024}:
 \begin{gather}
     C  = C_{LR}  = \frac{1}{2\pi} [W_y (\pi) - W_y (-\pi)], \label{eq:wy} 
 \end{gather}
 where $W_y$ measures the Berry phase in the $y$ direction:
 \begin{gather}
       W_y(k_x)  =  \int_{-\pi}^{\pi} dk_y \frac{d_x \partial_{k_y}d_y - d_y \partial_{k_y}d_x}{2d(d+d_z)}.
 \end{gather}

Let us first show that non-singular   exactly flat non-Bloch bands are topologically trivial. 
In \Cref{fig:chern}~(b), starting from the singular line $t_1^2 - \gamma_1^2 =  t_2^2 - \gamma_2^2$, we remove the singularity and open a gap  by tuning  $\gamma_i \to \gamma_i + \epsilon  \ (i = 1,2)$. Simultaneously, the flat band condition is kept with $t = -\sqrt{t_1^2 - \gamma_1^2}$. Varying $k_x$ across the real BZ, we observe there is net zero difference in the Berry phase.  It validates the no-go theorem with $C = 0$. Meanwhile, in the non-singular case, the non-Bloch torus Hamiltonian ($t_i^\pm \to t_i^\pm r^{\pm 1}$, $i=2,3$) can support on square plaquettes $N = (N_1 - 1)(N_2-1)$ independent CLSs [blue, \Cref{fig:chern}~(a)]. They constitute a complete and strictly local basis of the flat band, acting as the real-space obstruction.

Yet, there arise topological non-Bloch Chern bands by breaking exact band flatness.  We introduce a Hermitian perturbation $\delta H = \epsilon \sigma_y, \epsilon \in \mathbb{R}$ to the singular non-Bloch Hamiltonian $H_C$. Physically, it plays the role of an imaginary hopping term inside each unit cell and does not change its form under the similarity transformation when $\gamma_1 = 0, \delta r =1$, $r = \sqrt{\frac{t_2^-}{t_2^+}}$. In \Cref{fig:chern}~(c), perturbed by $\delta H$ with $\epsilon < 0$, a gap is opened. The lower band becomes nearly flat and acquires a nontrivial Chern number $C = -1$, as can be seen from a change of the Berry phase $\Delta W_y = -2\pi$. The upper band shares an opposite Chern number with the total Chern number vanishing: $C_+ + C_- = 0$. One can also analyze the topology from the effective \textcolor{black}{Hermitian} non-Bloch Hamiltonian around $\vec{k} = (0,0)$:
\begin{gather}
    \tilde{H}_C(\vec{k}) = \frac{k_y^2 - k_x^2}{2} \sigma_z - k_xk_y \sigma_x + \epsilon \sigma_y + \frac{k_x^2 + k_y^2}{2} \sigma_0.
\end{gather}
It is valid for $t_1 = -t = 1$, $\gamma_1 = 0$ and $t_2^+ t_2^- = t_1^2$. From the Berry curvature of the nearly flat band, one obtains
\begin{gather}
    C  = \int \frac{d^2k}{2\pi}  \frac{4\epsilon k^2}{(k^4 + 4\epsilon^2)^\frac{3}{2}} = \sgn(\epsilon),
\end{gather}
consistent with the results from the Berry phase. On the other hand, Hermitian perturbations $\delta H = \epsilon \sigma_x, \epsilon \sigma_z$ cannot open the gap along the singular line.

\clearpage

\begingroup
\onecolumngrid
\newpage
\appendix
\section{\large Supplementary Material for ``Non-Bloch Topology of Frustrated Non-Hermitian Flat Bands''}
In this supplementary material, we apply the Amoeba formulation to our class of frustrated NH flat band models. It provides an alternative proof for the exact uniform GBZ.

\renewcommand{\theequation}{S\arabic{equation}}
\setcounter{equation}{0} 

\subsection{Exact GBZ in the Amoeba formulation}
 In the Amoeba formulation~\cite{wang2024}, we first transform the Bloch Hamiltonians of NH breathing Kagome  and modified checkerboard models from $H(\vec{k})$ to $H(\vec{\beta})$ where  $e^{ik_l}$ is replaced by $\beta_l$ and $l=1,2$ or $l=x,y$. It renders
\begin{gather}
       H_K (\vec{\beta}) 
       =  \begin{pmatrix}
	  		0 & t_1^+ + t_2^- \beta_1^{-1} &    t_1^+ + t_2^-\beta_2^{-1} \\
			t_1^- + t_2^+\beta_1 & 0 & t+t\beta_1 \beta_2^{-1} \\
			 t_1^- + t_2^+ \beta_2 & t+t\beta_1^{-1} \beta_2 & 0
		    \end{pmatrix}, \notag \\
     H_C (\vec{\beta}) 
       =  \begin{pmatrix}
	  		-t(1-\beta_y t_2^+/t_1^-)(1-\beta_y^{-1}{t_2^-}/{t_1^+}) & t_1^+ (1-\beta_x^{-1})(1-\beta_y^{-1} {t_2^-}/{t_1^+})  \\
			 t_1^-(1-\beta_x)(1-\beta_yt_2^+/t_1^-)  & -t(1-\beta_x)(1-\beta_x^{-1})
		    \end{pmatrix}.       
	\end{gather}
In the GBZ, $|\beta_l| \ne 1$. The Amoeba is defined as a logarithmic map of the solutions to the characteristic equation of $H(\vec{\beta})$:
\begin{gather}
   \mathcal{A}_h = \{ \vec{\mu} \  \text{with} \   \mu_l = \log |\beta_l|: h(\vec{\beta}) = 0 \},
   \end{gather}
   where 
   \begin{gather}
   h (\vec{\beta}) = \det [H(\vec{\beta}) - E]/(E_{\text{FB}}-E), \label{eq:hfb}
\end{gather}
with $E$  the reference energy.
As the flat band (FB) is not affected by the GBZ shift, we remove its contribution to $h (\vec{\beta})$.  More explicitly,
\begin{gather}
    h_K(\vec{\beta}) = E^2 -  2tE - 2t^2 - t^2 (\beta_1\beta_2^{-1}+\beta_1^{-1}\beta_2) - t_1^-t_2^- (\beta_1^{-1} + \beta_2^{-1} ) - t_1^+t_2^+ (\beta_1 + \beta_2 ), \notag \\
       h_C(\vec{\beta}) = - E -t(1-\beta_yt_2^+/t_1^-)(1-\beta_y^{-1} {t_2^-}/{t_1^+}) -t(1-\beta_x)(1-\beta_x^{-1}).
\end{gather}

The Amoeba can be used to determine the GBZ and the boundary of the universal spectrum of NH Hamiltonians, which is the OBC spectrum  in the thermodynamic limit $E^{\text{OBC}}$. 
If $E$ is outside of $E^{\text{OBC}}$, the body of the Amoeba in the 2D plane displays a hole. As $E$ approaches and touches $E^{\text{OBC}}$, the hole shrinks and closes at the central point, which gives imaginary momentum shifts at the energy $E$. In \Cref{fig:amoeba}~(a), for NH breathing Kagome model, we choose $E$ from the non-Bloch spectrum and observe no hole in the Amoeba. We identify the central point at $\Gamma  = (\log |r|, \log |r|)$ where $r = \sqrt{\frac{t_1^-t_2^-}{t_1^+t_2^+}}$, consistent with the uniform GBZ in \Cref{eq:knb} from similarity transformation.
 \begin{figure}[!htb]
\centering
\includegraphics[width=0.5\columnwidth]{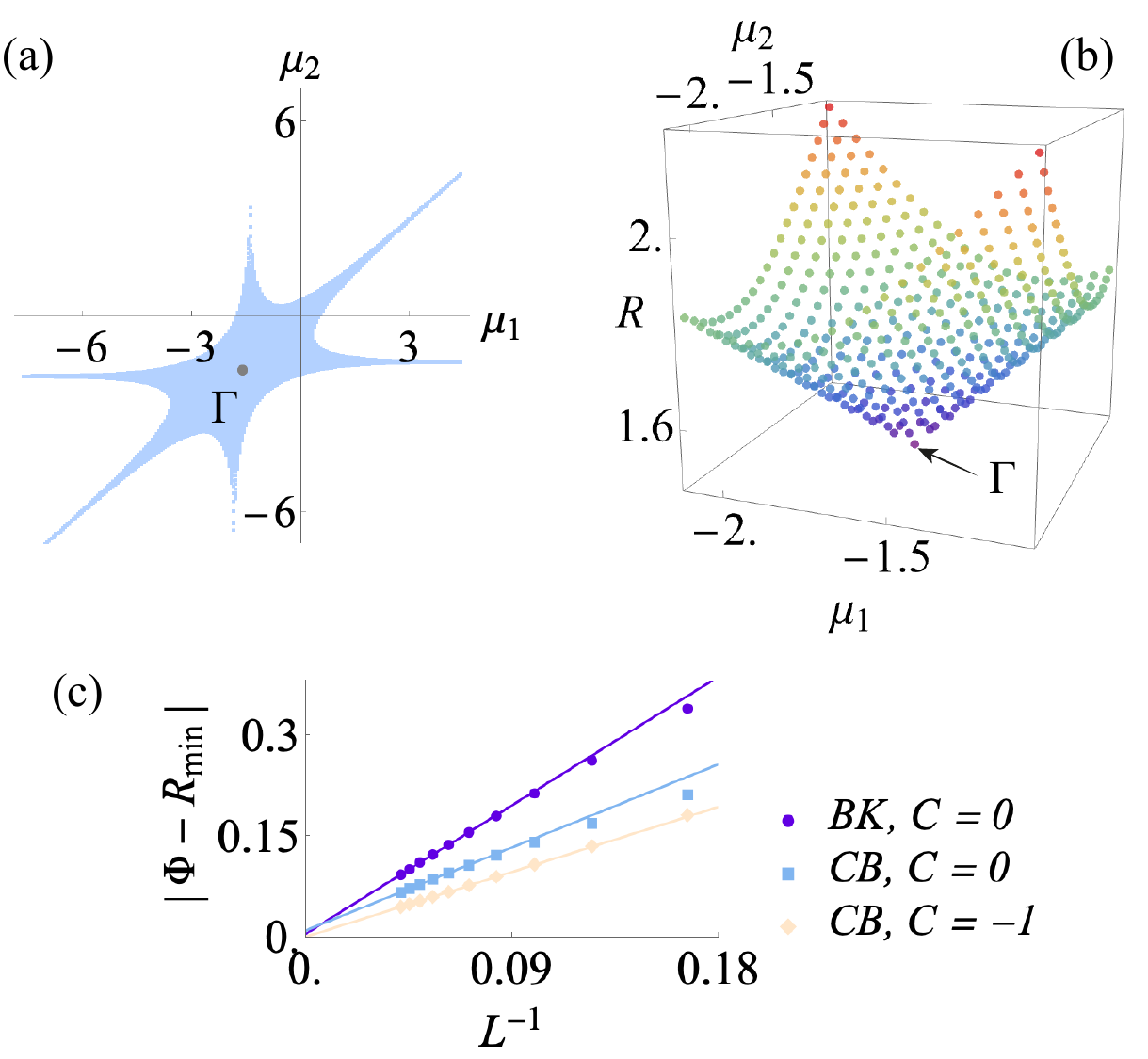}
\caption{{\bf \textcolor{black}{Exact GBZ in the Amoeba formulation}}. For the NH breathing kagome model in \Cref{eq:knb}, we realize the flat band condition with $t_1= 1, t_2=2$, $\gamma_1=1.5, \gamma_2 = 3$, $t = [(t_1^-t_1^++t_2^-t_2^+)/2]^{1/2}$. Given an arbitrary reference energy from the non-Bloch spectrum: $E = E_{-}(\frac{\pi}{5}, \frac{\pi}{5}) \simeq -1.43 i$, (a) the hole of the Amoeba closes  and (b) the Ronkin function  holds a single minimum at the central point $\Gamma = (\log |r|, \log |r|) \simeq (-1.61, -1.61)$ .  The observation that  $\Gamma$  is invariant under any choice of non-Bloch energies supports an exact uniform GBZ.   (c) Finite-size scaling of Coulomb potential difference based on the OBC spectrum and Ronkin function. We show the results of the NH breathing kagome (BK) and checkerboard (CB) models using geometries and parameters in \Cref{fig:sketch} and \Cref{fig:chern}. When the lowest band becomes exactly flat with $C = 0$ [\Cref{fig:sbt}~(b) and \Cref{fig:chern}~(b)], we evaluate the Coulomb potential from the other dispersive bands at $E = E_\text{FB}$. Whereas, when the lowest band becomes nearly flat with $C = -1$ [\Cref{fig:chern}~(c)], we examine the potential from all bands at $E = i$. The Ronkin function minimum is taken at $\Gamma = \vec{\mu}_c$ predicted by the exact GBZ. As the linear system size $L = N_1 = N_2$ grows, the OBC spectra of all three models converge to the universal spectra. The Ronkin functions in (b) and (c) are calculated on an integral grid with size $N_1 = N_2 = 100$.
}
\label{fig:amoeba}
\end{figure}
    
More rigorously, the central point of Amoeba $\Gamma = \vec{\mu}_c$ lies at the minimum of the Ronkin function [see \Cref{fig:amoeba}~(b)]:
\begin{gather}
   \left. \partial_{\mu_l} R_h(\vec{\mu}, E)\right|_{\vec{\mu} = \vec{\mu}_c} = 0, \quad \forall \mu_l,
\end{gather}
where the Ronkin function is given by
\begin{gather}
    R_h(\vec{\mu}, E) = \int_{T^2} \frac{\prod_l dq_l}{(2\pi)^2} \log |h(e^{\vec{\mu}+i\vec{q}})|.
\end{gather}
In $\log|h|$,  we express $\vec{\beta} = e^{\vec{\mu}+i\vec{q}}$ under the convention  $\beta_l = e^{\mu_l + iq_l}$. The integration has the domain of a two-dimensional torus $T^2 = [-\pi, \pi]^2$. Next, we show for $\forall t_i^\pm \in \mathbb{R}$, the uniform GBZ structures of our two models in \Cref{eq:knb} and \Cref{eq:cnbh} are exact. It is equivalent to prove
\begin{gather}
  \left.  \partial_{\mu_l} R_h(\vec{\mu}, E)\right|_{\vec{\mu} = \vec{\mu}_c} = 0, \quad \forall \mu_l, \forall E, \label{eq:rkm}
\end{gather}
where $\vec{\mu}_{c, K} = (\log r, \log r)$ and $\vec{\mu}_{c, C} = (0, \log r)$. First, we replace $\partial_{\mu_l}$ by $-i\partial_{q_l}$ and introduce the winding number $w_l$:
\begin{gather}
    \left. \partial_{\mu_l} R_h(\vec{\mu}, E)\right|_{\vec{\mu} = \vec{\mu}_c} = \int_{-\pi}^{\pi} \frac{dq_j}{2\pi} w_l (\vec{\mu}_c, E), \quad j \ne l, \notag \\
   w_l (\vec{\mu}_c, E) = \int_{-\pi}^{\pi} \frac{dq_l}{2\pi i} \frac{\partial_{q_l} h(e^{\vec{\mu}_c+i\vec{q}}) }{h(e^{\vec{\mu}_c+i\vec{q}})}.
\end{gather}

In NH breathing Kagome model, 
\begin{gather}
    \partial_{q_l} h_K (e^{\vec{\mu}_c+i\vec{q}}) = -f_l(\vec{q}) + f_l(-\vec{q}), 
\end{gather}
where $f_1(\vec{q}) = it^2 e^{i(q_1 - q_2)} +it_1^+t_2^+re^{iq_1}$ and $f_2(\vec{q}) = -it^2 e^{i(q_1 - q_2)} +it_1^+t_2^+re^{iq_2}$. Hence, 
\begin{gather}
 \forall \mu_l, \forall E: \quad  \left. \partial_{\mu_l} R_h(\vec{\mu}, E)\right|_{\vec{\mu} = \vec{\mu}_c}   = \int_{T^2} \frac{\prod_j dq_j}{(2\pi)^2} \frac{1}{i} \left( -\frac{f_l(\vec{q})}{h_K(\vec{\mu}_c, \vec{q})} + \frac{f_l(\vec{q})}{h_K(\vec{\mu}_c, -\vec{q})}  \right)  = 0.
\end{gather}
We have applied the emergent spectral symmetry  at $\vec{\mu}_{c,K} = (\log r, \log r)$:
\begin{gather}
    h_K( \vec{\mu}_c, \vec{q}) = h_K( \vec{\mu}_c, -\vec{q}).
\end{gather}

In NH modified checkerboard model,
\begin{gather}
    \partial_{q_l} h_C(e^{\vec{\mu}_c+i\vec{q}}) = p_l(q_l) - p_l (-q_l), 
\end{gather}
where $p_x(q_x) = i t e^{iq_x}$ and $p_y(q_y) = i t  (t_2^+r / t_1^-) e^{iq_y}$. It leads to vanishing winding numbers,
\begin{gather}
   \forall \mu_l, \forall E: \quad   w_l (\vec{\mu}_c, E) =\int_{-\pi}^{\pi} \frac{dq_l}{2\pi i} \left( \frac{p_l(q_l)}{h_C(\vec{\mu}_c, q_l, q_j)} - \frac{p_l(q_l)}{h_C(\vec{\mu}_c, -q_l,q_j)}  \right)  = 0.
\end{gather}
We also applied emergent spectral symmetries at $\vec{\mu}_{c,C} = (0, \log r)$:  
\begin{gather}
   h_C( \vec{\mu}_c, q_x, q_y) = h_C( \vec{\mu}_c, -q_x,q_y) = h_C( \vec{\mu}_c, q_x, -q_y).
\end{gather}
The proofs for \Cref{eq:rkm} are completed.

In the end, for dispersive bands in our models, we show their spectral convergence to the universal spectrum. In terms of electrostatics~\cite{lee2022d,wang2024}, we can assign electric charge $(-1/M)$ to each eigen energy $\epsilon_m$ on the dispersive bands with $M = N_1N_2$ the total number of unit cells. The total charge amounts to the number of dispersive bands. On the complex energy plane, the density of states $\rho$ at a reference energy $E$ can be obtained from the two-dimensional Coulomb potential $\Phi (E)$:
\begin{gather}
    \rho (E) = \frac{1}{2\pi} \Delta \Phi(E), \quad \Phi(E) = \frac{1}{M} \sum_{m \notin \text{FB}} \log |E - \epsilon_m|.
\end{gather}
Here, we take the Laplacian $\Delta = \frac{\partial^2}{\partial (\text{\re{E}})^2} + \frac{\partial^2}{\partial (\text{\im{E}})^2}$. In the thermodynamic limit $M \to \infty$, it turns out that the Coulomb potential is given by the Ronkin function minimum $R_{\min} (E)$, which can be resolved in our models with the exact GBZ [\Cref{eq:rkm}]:
\begin{gather}
  \lim_{M \to \infty} \Phi(E)  = \min_{\forall \vec{\mu}} R_h (\vec{\mu}, E) = R_h (\vec{\mu}_c, E).
\end{gather}
When $M$ is finite, we assume $N_1 = N_2 = L$ and $M = L^2$. As the linear system size $L$ increases, 
the Coulomb potential approaches the Ronkin function minimum  and their difference scales as~\cite{wang2024}
\begin{gather}
    \Phi(E) = R_{\min} (E) + O(L^{-1}).
\end{gather}
In \Cref{fig:amoeba}~(c), we plot the difference between Coulomb potential and Ronkin function minimum as a function of $L^{-1}$. We test the dispersive OBC spectra of NH breathing Kagome and modified checkerboard models in \Cref{fig:sbt}~(b), \Cref{fig:chern}~(b) and (c) using lattice geometries in \Cref{fig:sketch}~(a) and \Cref{fig:chern}~(a). In particular, in the case of \Cref{fig:chern}~(c), we modify $h(\vec{\beta})$ from \Cref{eq:hfb} to $h(\vec{\beta}) = \det [H(\vec{\beta}) - E]$ so that  the contribution from the nearly flat Chern band ($C = -1$) is also included. A finite-size scaling shows that for all three cases, the Coulomb potential difference follows  the $L^{-1}$ scaling and when extrapolated to $L \to \infty$, it converges to zero.

\endgroup

\end{document}